\documentclass{llncs}

\usepackage{paradise}
\usepackage{longtable,booktabs}
\usepackage{graphicx}
\usepackage[sectionbib]{natbib}
\setcitestyle{numbers,square,comma}

\usepackage{caption}
\begin{document}
\bibliographystyle{plainnat}

\pagestyle{plain}
\mainmatter

\title{From Model Checking to Runtime \break Verification and Back\thanks{This work has been partially supported by the Czech Science Foundation grant No. 15-08772S and by Red Hat, Inc.}}

\author{Katarína Kejstová \and Petr Ročkai \and Jiří Barnat}
\institute{\fimuni\\ \{xkejstov,xrockai,barnat\}@fi.muni.cz}

\maketitle

\begin{abstract}
  We describe a novel approach for adapting an existing software model checker to perform precise runtime verification. The software under test is allowed to communicate with the wider environment
  (including the file system and network). The modifications to the model checker are small and self-contained, making this a viable strategy for re-using existing model checking tools in a new context.
  
  Additionally, from the data that is gathered during a single execution in the runtime verification mode, we automatically re-construct a description of the execution environment which can then be used
  in the standard, full-blown model checker. This additional verification step can further improve coverage, especially in the case of parallel programs, without introducing substantial overhead into
  the process of runtime verification.
\end{abstract}

\section{Introduction}\label{introduction}

While model checking is a powerful technique for software verification, it also has certain limitations and deficiencies. Many of those limitations are related to the fact that a model checker must,
by design, fully isolate the program from any outside effects. Therefore, for verification purposes, the program under test is placed into an artificial environment, which gives non-deterministic (but
fully reproducible) responses to the program. The existence of this model environment immediately requires trade-offs to be made. If the environment model is too coarse, errors may be missed, or
spurious errors may be introduced. Creating a detailed model is, however, more costly, and the result is not guaranteed to exactly match the behaviour of the actual environment either. Moreover, a
detailed model may be too rigid: programs are often executed in conditions that have not been fully anticipated, and a certain amount of coarseness in the model of the environment can highlight such
unwarranted assumptions.

Many of those challenges are, however, not unique to model checking. In the context of automated testing, the test environment plays a prominent role, and a large body of work deals with related
problems. Unfortunately, adapting the methods used in automated testing to the context of model checking is far from straightforward. Making existing test-based setups easier to use with model
checking tools is a core contribution of this paper.

Both manual and automated testing are established, core techniques which play an important role in virtually every software development project. In a certain sense, then, testing provides an excellent
opportunity to integrate rigorous tools into the software development process. A number of verification tools specifically tailored for this mode of operation have seen great success in the software
development community, for instance the \texttt{memcheck} tool from the \texttt{valgrind} suite. We show that it is possible to tap into this potential also with a traditionally-designed software
model checker: we hope that this will help put powerful verification technology into the hands of software developers in a natural and seamless fashion. The second important contribution of this
paper, then, is an approach to build a runtime verification tool out of an existing software model checker.

Our main motivating application is extending our existing software model checker, \divine{}~\citep{barnat13:divine}, with a runtime verification mode. In its latest version, \divine{} has been split into a
number of well-defined, reusable components~\citep{rockai18:divm} and this presented an opportunity to explore the contexts in which the new components could be used. Based on this motivation, our
primary goal is to bring traditional (software) model checking and runtime verification closer together. As outlined above, there are two sides to this coin. One is to make model checking fit better
into existing software development practice, the second is to derive powerful runtime verification tools from existing model checkers. To ensure that the proposed approach is viable in practice, we
have built a prototype implementation, which allowed us to execute simple C and C++ programs in the resulting runtime verifier.

The rest of the paper is organised as follows: Section~\ref{sec:related} describes prior art and related work, while Section~\ref{sec:prelim} lays out our assumptions about the model checker and its
host environment. Section~\ref{sec:passthrough} describes adapting a model checker to also work as a runtime verifier and Section~\ref{sec:replay} focuses on how to make use of data gathered by the
runtime verifier in the context of model checking. Section~\ref{sec:implementation} describes our prototype implementation based on \divine{} (including evaluation) and finally,
Section~\ref{sec:conclusion} summarises and concludes the paper.

\section{Related Work}\label{sec:related}

There are two basic approaches to runtime verification~\citep{havelund04:efficien}: online (real time) monitoring, where the program is annotated and, during execution, reports its actions to a
monitor. In an offline mode, the trace is simply collected for later analysis. Clearly, an online-capable tool can also work in offline mode, but the reverse is not always true. An extension of the
online approach allows the program to be monitored also in production, and property violations can invoke a recovery procedure in the program~\citep{meredith12:mop}. Our work, in principle, leads to
an online verifier, albeit with comparatively high execution overhead, which makes it, in most cases, unsuitable for executing code in production environments. Depending on the model checker used, it
can, however, report violations to the program and invoke recovery procedures and may therefore be employed this way in certain special cases.

Since our approach leads to a runtime verification tool, this can be compared to other such existing tools. With the exception of \texttt{valgrind}~\citep{nethercote07:valgrin}, most tools in this
category focus on Java programs. For instance, Java PathExplorer~\citep{havelund04:overview.runtim} executes annotated Java byte code, along with a monitor which can check various properties,
including past-time LTL. Other Java-based tools include JavaMOP~\citep{jin12:javamop} with focus on continuous monitoring and error recovery and Java-MaC~\citep{kim04:java.mac} with focus on
high-level, formal property specification.

Our \emph{replay mode} (described in Section~\ref{sec:replay}) is also related to the approach described in~\citep{havelund00:using.runtim}, where data collected at runtime is used to guide the model
checker, with the aim of reducing the size of the state space. In our case, the primary motivation is to use the model checker for verifying more complex properties (including LTL) and to improve
coverage of runtime verification.

\section{Preliminaries}\label{sec:prelim}

There are a few assumptions that we need to make about the mode of operation of the model checker. First, the model checker must be able to restrict the exploration to a single execution of the
program, and it must support explicitly-valued operations. The simplest case is when the model checker in question is based on an explicit-state approach (we will deal with symbolic and/or abstract
values in Section~\ref{sec:abstract}). If all values are represented explicitly in the model checker, exploration of a single execution is, in a sense, equivalent to simply running the program under
test. Of course, since this is a model checker, the execution is subject to strict error checking.

\subsection{Abstract and Symbolic Values}\label{sec:abstract}

The limitation to exploring only a single execution is, basically, a limitation on \emph{control flow}, not on the representation of variables. The root cause for the requirement of exploring only one
control flow path is that we need to insert actions into the process of model checking that will have consequences in the outside world, consequences which cannot be undone or replayed. Therefore, it
is not viable to restore prior states and explore different paths through the control flow graph, which is what normally happens in a model checker. It is, however, permissible to represent data in an
abstract or symbolic form, which essentially means the resulting runtime verifier will also act as a symbolic executor. In this case, an additional requirement is that the values that reach the
outside world are all concrete (the abstract representation used in the model checker would not be understood by the host operating system or the wider environment). Luckily, most tools with support
for symbolic values already possess this capability, since it is useful in a number of other contexts.

\subsection{Environments in Model Checking}\label{sec:env}

A model checker needs a complete description of a system, that is, including any environment effects. This environment typically takes the form of code in the same language as the program itself, in
our case C or C++. For small programs or program fragments, it is often sufficient to write a custom environment from scratch. This is analogous to how unit tests are written: effects from outside of
the program are captured by the programmer and included as part of the test.

When dealing with larger programs or subsystems, however, the environment becomes a lot more complicated. When the program refers to an undefined function, the model checker will often provide a
fallback implementation that gives completely undetermined results. This fallback, typically, does not produce any side effects. Such fallback functions constitute a form of synthetic model
environment. However, this can be overly coarse: such model environment will admit many behaviours that are not actually possible in the real one, and vice versa, lasting side effects of a program
action (for instance a change in file content) may not be captured at all. Those infidelities can introduce both false positives and false negatives. For this reason, it is often important to provide
a more realistic environment.

A typical model checker (as opposed to a runtime verifier) cannot make use of a real operating system nor of testing-tailored, controlled environment built out of standard components (physical or
virtual machines, commodity operating systems, network equipment and so on). A possible compromise is to implement an operating system which is designed to run inside a model checker, as a stand-in
for the real OS. This operating system can then be re-used many times when constructing environments for model checking purposes. Moreover, this operating system is, to a certain degree, independent
of the particular model checker in use. Like with standard operating systems, a substantial part of the code base can be re-used when porting the OS (that is, the host model checker is akin to a
processor architecture or a hardware platform in standard operating systems).

Many programs of interest are designed to run on POSIX-like operating systems, and therefore, POSIX interfaces, along with the interfaces mandated by ISO C and C++ are a good candidate for
implementation. This has the additional benefit that large parts of all these specifications are implemented in open source C and/or C++ code, and again, large parts of this code are easily ported to
new kernels. Together, this means that a prefabricated environment with POSIX-like semantics is both useful for verifying many programs and relatively simple to create.

In the context of a model checker, the kernel of the operating system can be linked directly to the program, as if it were a library. In this approach, the model checker in question does not need any
special support for loading kernel-like objects or even for privilege separation.

\subsection{System Calls}\label{sec:syscall}

In this section, we will consider how traditional operating systems, particularly in the POSIX family, define and implement system calls. A traditional operating system consists of many different
parts, but in our context, the most important are the kernel and the user-space libraries which implement the operating system API (the most important of these libraries is, on a typical Unix system,
\texttt{libc}). From the point of view of a user program, the \texttt{libc} API \emph{is} the interface of the operating system. However, many functions which are mandated as part of this interface
cannot be entirely implemented in the user space: they work with resources that the user-space code is unable to directly access. Examples of such functions would be \texttt{read} or \texttt{write}:
consider a \texttt{read} from a file on a local file system. If the implementation was done in the user space, it would need direct access to the hardware, for instance the PCI bus, in order to talk
to the hard drive which contains the requisite blocks of data which represent the file system. This is, quite clearly, undesirable, since granting such access to the user program would make access
control and resource multiplexing impossible.

For these reasons, it is standard practice to implement parts of this functionality in separate, system-level code with a restricted interface, which makes access control and resource sharing
possible. In operating system designs with monolithic kernels, this restricted interface consists of what is commonly known as system calls.\footnote{In microkernel and other design schools, syscalls
  in the traditional sense only exist as an abstraction, and are implemented through some form of inter-process communication.} A system call is, then, a mechanism which allows the user-space code to
request that the system-level software (the kernel) executes certain actions on behalf of the program (subject to appropriate permission and consistency checks). The actual implementation of syscall
invocation is platform-specific, but it always involves a switch from user (non-privileged) mode into kernel mode (privileged mode, \emph{supervisor} mode or \emph{ring 0} on x86-style processors).

On POSIX-like systems, \texttt{libc} commonly provides a generic \texttt{syscall} function (it first appeared in \texttt{3BSD}). This function allows the application to issue syscalls based on their
number, passing arguments via an ellipsis (i.e.~by taking advantage of variadic arguments in the C calling convention). In particular, this means that given a description of a system call (its number
and the number and types of its arguments), it is possible to automatically construct an appropriate invocation of the \texttt{syscall} function.

\subsection{Overview of Proposed Extensions}\label{sec:overview}

Under the proposed extensions, we have a model checker which can operate in two modes: \emph{run} and \emph{verify}. In the \emph{run} mode, a single execution of the program is explored, in the
standard execution order. We expect that all behaviour checking (enforcement of memory safety, assertion checks, etc.) is still performed in this mode. The \emph{verify} mode, on the other hand, uses
the standard model checking algorithm of the given tool.

\begin{figure}
\centering
\includegraphics{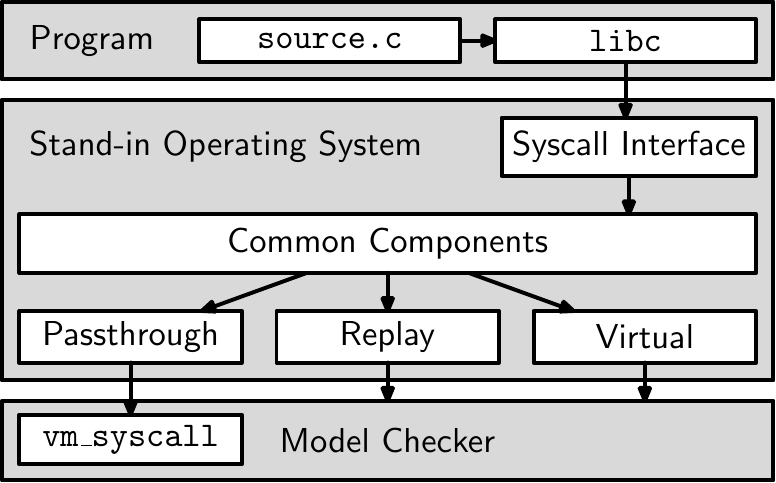}
\caption{A scheme of components involved in our proposed approach.}\label{fig:scheme}
\end{figure}

The system under test (the input to this model checker), then, consists of the user program itself, along with the environment, the latter of which contains a stand-in operating system. The situation
is illustrated in Figure~\ref{fig:scheme}. The operating system has 3 different modes:

\begin{enumerate}
\def\labelenumi{\arabic{enumi}.}
\tightlist
\item
  a \emph{virtual} mode, in which all interaction with the real world is simply simulated -- for example, a virtual file system is maintained in-memory and is therefore part of the state of the system
  under test; this OS mode can be used with both \emph{run} and \emph{verify} modes of the model checker
\item
  a \emph{passthrough} mode, which uses the \texttt{vm\_syscall} model checker extension to execute system calls in the host operating system and stores a trace of all the syscalls it executed for
  future reference; this OS mode can only be used in the \emph{run} mode of the model checker
\item
  a \emph{replay} mode, which reads the system call trace recorded in the \emph{passthrough} mode, but does not interact with the host operating system; this OS mode can be again used in both the
  \emph{run} and \emph{verify} mode of the model checker
\end{enumerate}

\section{Syscall Passthrough}\label{sec:passthrough}

In order to turn a model checker into a runtime verifier, we propose a mechanism which we call \emph{syscall passthrough}, where the virtual, stand-in operating system (see Section~\ref{sec:env})
gains the ability to execute syscalls in the host operating system (see also Section~\ref{sec:syscall}). Of course, this is generally \emph{unsafe}, and only makes sense if the model checker can
explore a single run of the program and do so \emph{in order}.

Thanks to the architecture of system calls in POSIX-like kernels, we only need a single new primitive function to be implemented in the model checker (we will call this new primitive function
\texttt{vm\_syscall} from now on; first, we need to avoid confusion with the POSIX function \texttt{syscall}, second, the model checker acts as a virtual machine in this context). The sole purpose of
the function is to construct and execute, in the context of the host operating system, an appropriate call to the host \texttt{syscall} function (the interface of which is explained in more detail in
Section~\ref{sec:syscall}).

We would certainly like to avoid any system-specific knowledge in the implementation of \texttt{vm\_syscall} -- instead, any system-specific code should reside in the stand-in OS, which is much easier
to modify than the model checker proper. To this end, the arguments to our \texttt{vm\_syscall} primitive contain metadata describing the arguments \texttt{syscall} expects, in addition to the data
itself. That is, \texttt{vm\_syscall} needs to know whether a particular argument is an input or an output argument, its size, and if it is a buffer, the size of that buffer. The exact encoding of
these metadata will be described in Section~\ref{sec:passthrough-mc}, along with more detailed rationale for this approach.

Finally, most of the implementation work is done in the context of the (stand-in) operating system (this is described in more detail in Section~\ref{sec:passthrough-os}). This is good news, because
most of the code in the operating system, including all of the code related to syscall passthrough, is in principle portable between model checkers.

\subsection{Model Checker Extension}\label{sec:passthrough-mc}

The model checker, on the other hand, only needs to provide one additional primitive. As already mentioned, we call this primitive \texttt{vm\_syscall}, and it should be available as a variadic C
function to the system under test. This is similar to other built-in functions often provided by model checkers, like \texttt{malloc} or a non-deterministic choice operator. While in the program under
test, invocations of such built-ins look just like ordinary C function calls, they are handled differently in the model checker and often cause special behaviour that is not otherwise available to a C
program.

We would like this extension to be as platform-neutral as possible, while maintaining simplicity. Of course not all platforms provide the \texttt{syscall} primitive described in
Section~\ref{sec:syscall}, and on these platforms, the extension will be a little more complicated. Namely, when porting to a platform of this type, we need to provide our own implementation of
\texttt{syscall}, which is easy to do when the system calls are available as C functions, even if tedious. In this case, we can simply assign numbers to system calls and construct a single
\texttt{switch} statement which, based on a number, calls the appropriate C function.

Therefore, we can rely on the \texttt{syscall} system-level primitive without substantial loss of generality or portability. The next question to ask is whether a different extension would serve our
purpose better -- in particular, there is the obvious choice of exposing each syscall separately as a model checker primitive. There are two arguments against this approach.

First, it is desirable that the syscall-related machinery is all in one place and not duplicated in both the stand-in operating system and in the model checker. However, in the \emph{virtual} and
\emph{replay} modes, this machinery must be part of the stand-in operating system, which suggests that this should be also the case in the \emph{passthrough} mode.

Second, the number of system calls is quite large (typically a few hundred functions) and the functions are system-dependent. When the code that is specific to the host operating system resides in the
stand-in operating system, it can be ported once and multiple model checkers can benefit. Of course, the stand-in operating system needs to be ported to the model checker in question, but this offers
many additional advantages (particularly the virtual mode).

Now if we decide that a single universal primitive becomes part of the model checker, we still need to decide the syntax and the semantics of this extension. Since different system calls take
different arguments with varying meaning, the primitive itself will clearly need to be variadic. Since one of the main reasons for choosing a single-primitive interface was platform neutrality, the
primitive itself should not possess special knowledge about individual syscalls. First of all, it does not know the bit widths of individual arguments (on most systems, some arguments can be 32 bit --
for instance file descriptors -- and other 64 bit -- object sizes, pointers, etc.). This information is crucial to correctly set up the call to \texttt{syscall} (the variadic arguments must line up).
Moreover, some pointer-type arguments represent variable-sized \emph{input} data (the buffer argument to \texttt{write}, for example) and others represent \emph{output} data (the buffer argument to
\texttt{read}). In both cases, the size of the memory allocated for the variable-sized argument must be known to \texttt{vm\_syscall}, so that this memory can be correctly copied between the model
checker and the system under test.

\begin{figure}
\centering
\includegraphics{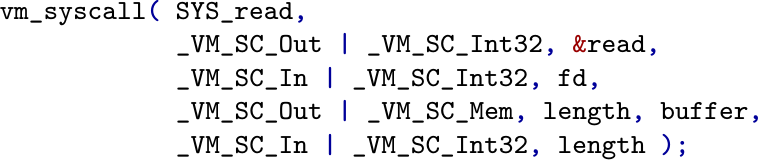}
\caption{An example invocation of \texttt{vm\_syscall} performing a \texttt{read} passthrough.}\label{fig:syscall}
\end{figure}

For these reasons, the arguments to \texttt{vm\_syscall} also contain metadata: for each real argument that ought to be passed on to \texttt{syscall}, 2 or 3 arguments are passed to
\texttt{vm\_syscall}. The first one is always type information: whether the following argument is a scalar (32b or 64b integer) or a pointer, whether it is an input or an output. If the value is a
scalar input, the second argument is the value itself, if it is a scalar output, the following argument is a pointer to an appropriate-sized piece of memory. If the value is a pointer, the size of the
pointed-to object comes second and the pointer itself comes third. An example invocation of \texttt{vm\_syscall} is shown in Figure~\ref{fig:syscall}. The information passed to \texttt{vm\_syscall}
this way is sufficient to both construct a valid call to \texttt{syscall} and to copy inputs from the system under test to the host system and pass back the outputs.

\subsection{Operating System Extension}\label{sec:passthrough-os}

The \texttt{vm\_syscall} interface described above is a good low-level interface to pass syscalls through to the host operating system, but it is very different from the usual POSIX way to invoke
them, and it is not very intuitive or user-friendly either. It is also an unsafe interface, because wrong metadata passed to \texttt{vm\_syscall} can crash the model checker, or corrupt its memory.

The proper POSIX interface is to provide a separate C function for each syscall, essentially a thin wrapper that just passes the arguments along. Calling these dedicated wrappers is more convenient,
and since they are standard C functions, their use can be type-checked by the compiler. In the \emph{virtual} mode of the operating system, those wrappers cause the execution to divert into the
kernel. We can therefore re-use the entire \texttt{libc} without modifications, and implement syscall passthrough at the kernel level, where we have more control over the code.

In our OS design, the kernel implements each system call as a single C++ method of a certain class (a \emph{component}). Which exact components are activated is decided at boot time, and it is
permissible that a given system call is implemented in multiple components. Since the components are arranged in a stack, the topmost component with an implementation of a given system call ``wins''.
In this system, implementing a passthrough mode is simply a question of implementing a suitable passthrough component and setting it up. When \texttt{libc} invokes a system call, the control is
redirected into the kernel as usual, and the passthrough component can construct an appropriate invocation of \texttt{vm\_syscall}.

This construction requires the knowledge of a particular system call. Those are, luckily, more or less standardised by POSIX and the basic set is therefore reasonably portable. Moreover, we already
need all of this knowledge in the implementation of the virtual mode, and hence most of the code related to the details of argument passing can be shared. As mentioned earlier, this means that the
relevant \texttt{libc} code and the syscall mechanism it uses internally is identical in all the different modes of operation. The passthrough mode is, therefore, implemented entirely in the kernel of
the stand-in operating system.

\subsection{Tracing the Syscalls}\label{sec:tracing}

The architecture of syscall passthrough makes it easy to capture argument values and results of every invoked syscall, in addition to actually passing it on to the host operating system. Namely, the
implementation knows exactly which arguments are inputs and which are outputs and knows the exact size of any buffer or any other argument passed as a pointer (both input and output). This allows the
implementation to store all this data in a file (appending new records as they happen). This file can then be directly loaded for use in the \emph{replay mode} of the stand-in operating system.

\section{Syscall Replay}\label{sec:replay}

In a model checker, all aspects of program execution are fully repeatable. This property is carried over into the \emph{virtual} operating mode (as described in this paper), but not into the
\emph{passthrough} mode. System calls in the host operating system are, in general, not repeatable: files appear and disappear and change content, network resources come and go and so on, often
independently of the execution of the program of interest.

What the passthrough mode can do, however, is recording the answers from the host operating system (see Section~\ref{sec:tracing}). When we wish to repeat the same execution of the program (recall
that everything apart from the values coming from \texttt{vm\_syscall} is under the full control of the model checker), we do not need to actually pass on the syscalls to the host operating system:
instead, we can read off the outputs from a trace. This is achieved by simply replacing all invocations of \texttt{vm\_syscall} by a different mechanism, which we will call \texttt{replay\_syscall}.
This new function looks at the trace, ensures that the syscall invoked by the program matches the one that comes next in the trace and then simply plays back the effects observable in the program.
Since the program is otherwise isolated by the model checker, those effects are limited to the changes the syscall caused in its output parameters and the value of \texttt{errno}. The appropriate
return value is likewise obtained from the trace.

\subsection{Motivation}\label{motivation}

There are two important applications of the replay mode. First, if the model checker in question provides interactive tools to work with the state space, we can use those tools to look at real
executions of the program, and in particular, we can easily step backwards in time. That is, if we have an interactive simulator (like, for example, presented in~\citep{rockai17:simulat.llvm.bitcod}),
we can derive a reversible debugger essentially for free by recording an execution in the passthrough mode and then exploring the corresponding path through the state space in the \emph{replay} mode.

Second, if the behaviour of the program depends on circumstances other than the effects and return values of system calls, it is often the case that multiple different executions of the program will
result in an identical sequence of system calls. As an example, if the program contains multiple threads, one of which issues syscalls and others only participate in computation and synchronisation,
the exact thread interleaving will only have a limited effect on the order and arguments of system calls, if any. The model checker is free to explore all such interleavings, as long as they produce
the same syscall trace.

That this is a practical ability is easily demonstrated. A common problem is that a given program, when executed in a controlled environment, sometimes executes correctly and other times incorrectly.
In this case, by a controlled environment we mean that files and network resources did not change, and that the behaviour of the program does not depend on the value of the real-time clock. Therefore,
we can reasonably expect the syscall trace to be identical (at least up to the point where the unexpected behaviour is encountered). If this is the case, the model checker will be able to reliably
detect the problem based on a single syscall trace, regardless of whether the problem did or did not appear while running in the passthrough mode.

\subsection{Constructing the State Space}\label{constructing-the-state-space}

As explained above, we can use the replay mode to explore behaviours of the program that result in an identical syscall trace, but are not, computation-wise, identical to the original passthrough
execution. In this case, it is important that the model checker explores only executions with this property. A primitive which is commonly available in model checkers and which can serve this purpose
is typically known as \texttt{assume}\footnote{The \texttt{assume} primitive is a counterpart to \texttt{assert} and has a similar interface. It is customary that a single boolean value is given as a
  parameter to the \texttt{assume} statement (function call), representing the assumed condition.}. The effect of this primitive is to instruct the model checker to abandon any executions where the
condition of the \texttt{assume} does not hold. Therefore, our \texttt{replay\_syscall}, whenever it detects a mismatch between the syscall issued by the program and the one that is next in the trace,
it can simply issue \texttt{assume(\ false\ )}. The execution is abandoned and the model checker is forced to explore only those runs that match the external behaviour of the original.

\subsection{Causality-Induced Partial Order}\label{causality-induced-partial-order}

The requirement that the traces exactly match up is often unnecessarily constraining. For instance, it is quite obvious that the order of two read operations (with no intervening write operations) can
be flipped without affecting the outcome of either of the two reads. In this sense, such two reads are not actually ordered in the trace. This means that the trace does not need to be ordered linearly
-- the two reads are, instead, incomparable in the causal ordering. In general, it is impossible to find the exact causal relationships between syscalls, especially from the trace alone -- a write to
a file may or may not have caused certain bytes to appear on the \texttt{stdin} of the program. We can, however, construct an approximation of the correct partial order, and we can do so safely: the
constructed ordering will always respect causality, but it may order certain actions unnecessarily strictly.

We say that two actions \(a\) and \(b\) (system call invocations) \emph{commute} if the outcome of both is the same, regardless of their relative ordering (both \(a\) and \(b\) have the same
individual effect, whether they are executed as \(a, b\) or as \(b, a\)). Given a sequence of system calls that respects the causal relationships, swapping two adjacent entries which commute will lead
to a new sequence with the same property. We can obtain an approximate partial order by constructing all such sequences and declaring that \(a < b\) iff this is the case in all of the generated
sequences.

\section{Prototype Implementation}\label{sec:implementation}

\begin{figure}
\centering
\includegraphics{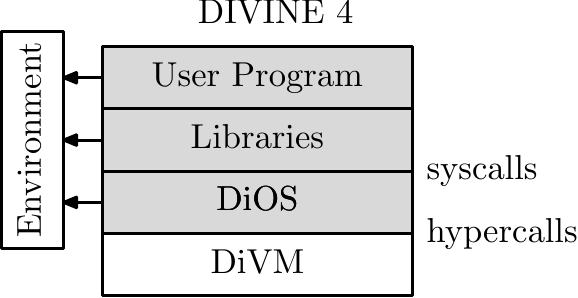}
\caption{The architecture of \divine{} 4. The shaded part is, from a model checking point of view, the system under test. However, \dios{} and most of the libraries are shipped as part of
\divine{}.}\label{fig:d4}
\end{figure}

We have implemented the approach described in this paper, using the \divine{} 4 software model checker as a base. In particular, we rely on the \divm{} component in \divine{}, which is a verification-focused
virtual machine based on the \llvm{} intermediate representation (more details in Section~\ref{sec:llvm}). The architecture of \divine{} 4, as a model checker, is illustrated in Figure~\ref{fig:d4}. First,
we have extended \divm{} with the \texttt{vm\_syscall} primitive (cf. Section~\ref{sec:passthrough}). Taking advantage of this extension, we have implemented the requisite support code in \dios{}, as
described in Section~\ref{sec:passthrough-os}. \dios{} is a pre-existing stand-in operating system component which originally supported only the \emph{virtual} mode of operation. As part of the work
presented in this paper, we implemented both a passthrough and a replay mode in \dios{}.

In the rest of this section, we will describe the underpinnings of \divine{} 4 in more detail. The first important observation is that, since \divine{} is based on interpreting \llvm{} bitcode, it can use a
standard compiler front-end to compile C and C++ programs into the bitcode form, which can then be directly verified. We will also discuss the limitations of the current implementation and demonstrate
its viability using a few examples.

\subsection{\llvm{} Bitcode}\label{sec:llvm}

\llvm{} bitcode (or intermediate representation)~\citep{llvm16:llvm.languag} is an assembly-like language primarily aimed at optimisation and analysis. The idea is that \llvm{}-based analysis and
optimisation code can be shared by many different compilers: a compiler front end builds simple \llvm{} IR corresponding to its input and delegates all further optimisation and native code generation to
a common back end. This architecture is quite common in other compilers: as an example, GCC contains a number of different front ends that share infrastructure and code generation. The major
innovation of \llvm{} is that the language on which all the common middle and back end code operates is exposed and available to 3rd-party tools. It is also quite well-documented and \llvm{} provides
stand-alone tools to work with both bitcode and textual form of this intermediate representation.

From a language viewpoint, \llvm{} IR is in partial SSA form (single static assignment) with explicit basic blocks. Each basic block is made up of instructions, the last of which is a \emph{terminator}.
The terminator instruction encodes relationships between basic blocks, which form an explicit control flow graph. An example of a terminator instruction would be a conditional or an unconditional
branch or a \texttt{ret}. Such instructions either transfer control to another basic block of the same function or stop execution of the function altogether.

Besides explicit control flow, \llvm{} also strives to make much of the data flow explicit, taking advantage of partial SSA for this reason. It is, in general, impossible to convert entire programs to a
full SSA form; however, especially within a single function, it is possible to convert a significant portion of code. The SSA-form values are called \emph{registers} in \llvm{} and only a few
instructions can ``lift'' values from memory into registers and put them back again (most importantly \texttt{load} and \texttt{store}, respectively, plus a handful of atomic memory access
instructions).

\subsection{Runtime Verification with \llvm{}}\label{runtime-verification-with-llvm}

While \llvm{} bitcode is primarily designed to be transformed and compiled to native code, it can be, in principle, executed directly. Of course, this is less convenient than working with native code,
but since the bitcode is appreciably more abstract than typical processor-level code, it is more amenable to model checking. The situation can be improved by providing tools which can work with hybrid
object files, which contain both native code and the corresponding \llvm{} bitcode. This way, the same binary can be both executed natively and analysed by \llvm{}-based tools.

\subsection{\llvm{} Extensions for Verification}\label{sec:extensions}

Unfortunately, \llvm{} bitcode alone is not sufficiently expressive to describe real programs: most importantly, it is not possible to encode interaction with the operating system into \llvm{} instructions.
When \llvm{} is used as an intermediate step in a compiler, the lowest level of the user side of the system call mechanism is usually provided as an external, platform-specific function with a standard C
calling convention. This function is usually implemented in the platform's assembly language. The system call interface, in turn, serves as a gateway between the program and the operating system,
unlocking OS-specific functionality to the program. An important point is that the gateway function itself cannot be implemented in portable \llvm{}.

To tackle these problems, a small set of primitives was proposed in \citep{rockai18:divm} (henceforth, we will refer to this enriched language as \divm{}). With these primitives, it is possible to
implement a small, isolated operating system in the \divm{} language alone. \divine{} already provides such an operating system, called \dios{} -- the core OS is about 2500 lines of C++, with additional 5000
lines of code providing \emph{virtual} POSIX-compatible file system and socket interfaces. Our implementation of the ideas outlined in Section~\ref{sec:passthrough-os} can, therefore, re-use a
substantial part of the existing code of \dios{}.

\subsection{Source Code}\label{source-code}

The implementation consists of two parts. The model checker extension is about 200 lines of C++, some of which is quite straightforward. The \dios{} extension is more complex: the passthrough component
is about 1400 lines, while the replay component is less than 600. All the relevant source code, including the entire \divine{} 4 model checker, can be obtained online\footnote{\url{https://divine.fi.muni.cz/2017/passthrough/}}.

\subsection{Limitations}\label{sec:limitations}

There are two main limitations in our current implementation. The first is caused by a simplistic implementation of the \emph{run} mode of our model checker (see Section~\ref{sec:overview}). The main
drawback of such a simple implementation is that syscalls that block may cause the entire model checker to deadlock. Specifically, this could happen in cases where one program thread is waiting for an
action performed by another program thread. Since there is only a single model checker thread executing everything, if it becomes blocked, no program threads can make any progress. There are two
possible counter-measures: one is to convert all system calls to non-blocking when corresponding \texttt{vm\_syscall} invocations are constructed, another is to create multiple threads in the model
checker, perhaps even a new thread for each system call. Only the latter approach requires additional modifications to the model checker, but both require modifications to the stand-in operating
system.

The second limitation stems from the fact that our current \texttt{libc} implementation only covers a subset of POSIX. For instance, the \texttt{gethostbyname} interface (that is, the component of
\texttt{libc} known as a resolver) is not available. This omission unfortunately prevents many interesting programs from working at the moment. However, this is not a serious limitation in principle,
since the resolver component from an existing \texttt{libc} can be ported. Many of the networking-related interfaces are already present and work (in particular, TCP/IP client functionality has been
tested, cf. Section~\ref{sec:evaluation}).

Finally, a combination of both those limitations means that the \texttt{fork} system call, which would create a new process, is not available. In addition to problems with blocking calls, there are a
few attributes that are allocated to each process, and those attributes can be observed by certain system calls. For example, one such attribute is the \texttt{pid} (process identifier), obtainable
with a \texttt{getpid} system call, another is the working directory of the process, available through \texttt{getcwd}. Again, there are multiple ways to resolve this problem, some of which require
modifications in the model checker.

\subsection{Evaluation}\label{sec:evaluation}

Mainly due to the limitations outlined in Section~\ref{sec:limitations}, it is not yet possible to use our prototype with many complete, real-world programs. The domain in which \divine{} has been mainly
used so far are either small, self-contained programs and unit tests for algorithms and data structures. Both sequential and parallel programs can be verified. The source distribution of \divine{}
includes about 600 test cases for the model checker, many of which also use POSIX interfaces, leveraging the existing \emph{virtual} mode of \dios{}. As a first part of our evaluation, we took all those
test cases and executed them in the new \emph{passthrough} mode, that is, in a mode when \divine{} acts as a runtime verifier. A total of 595 tests passed without any problems, 3 timed out due to use of
blocking system calls and 9 timed out due to presence of infinite loops. Of course, since runtime verification is not exhaustive, not all errors present in the 595 tests were uncovered in this mode.

The second part of our evaluation was to write small programs that specifically test the \emph{passthrough} and the \emph{replay} mode:

\begin{itemize}
\tightlist
\item
  \texttt{pipe}, which creates a named pipe and two threads, one writer and one reader and checks that data is transmitted through the pipe
\item
  \texttt{rw} which simply creates, writes to and reads from files
\item
  \texttt{rw-par} in which one thread writes data to a file and another reads and checks that data
\item
  \texttt{network}, a very simple HTTP client which opens a TCP/IP connection to a fixed IP address, performs an HTTP request and prints the result
\end{itemize}

We tested these programs in both the \emph{passthrough} mode and in the \emph{replay} mode. While very simple, they clearly demonstrate that the approach works. The source code of those test programs
is also available online\footnote{\url{https://divine.fi.muni.cz/2017/passthrough/}}. Clearly, our verifier incurs appreciable overhead, since it interprets the program, instead of executing it
directly. Quantitative assessment of the runtime and memory overhead is subject to future work (more complex test cases are required).

\section{Conclusions and Future Work}\label{sec:conclusion}

We have described an approach which allows us to take advantage of an existing software model checking tool in the context of runtime verification. On one hand, this approach makes model checking more
useful by making it usable with real environments while retaining many of its advantages over testing. On the other hand, it makes existing model checking tools useful in cases when runtime
verification is the favoured approach. The approach is lightweight, since the modification to the model checker is small and self-contained. The other component required in our approach, the stand-in
operating system, is also reasonably portable between model checkers. The overall effort associated with our approach is small, compared to implementing two dedicated tools (a model checker and a
runtime verifier).

In the future, we plan to remove the limitations described in Section~\ref{sec:limitations} and offer a production-ready implementation of both a passthrough and a replay mode in \divine{} 4. Since the
results of the preliminary evaluation are highly encouraging, we firmly believe that a runtime verification mode based on the ideas laid out in this paper will be fully integrated into a future
release of \divine{}.

\bibliography{common}

\end{document}